\documentclass[twoside,leqno,twocolumn]{article}
\usepackage{ltexpprt}

\usepackage{paralist}
\usepackage{algorithm}
\usepackage{algorithmic}
\usepackage{epsfig}
\usepackage{epstopdf}
\usepackage{subfigure}

\usepackage{amssymb}

\usepackage{amsmath}
\usepackage{multirow}
\usepackage{graphicx}

\usepackage{caption2}

\begin{document}

\renewcommand{\algorithmicrequire}{\textbf{INPUT:}}
\renewcommand{\algorithmicensure}{\textbf{OUTPUT:}}

\title{\Large Predicting Neighbor Distribution in Heterogeneous Information Networks}

\author{Yuchi Ma\thanks{School of Computer Science, Sichuan University, Chengdu, China. scu.Richard.Ma@gmail.com} \\
\and
Ning Yang\thanks{Corresponding author. School of Computer Science, Sichuan University, Chengdu, China. yangning@scu.edu.cn}\\
\and
Chuan Li\thanks{School of Computer Science, Sichuan University, Chengdu, China. lcharles@scu.edu.cn}\\
\and
Lei Zhang\thanks{School of Computer Science, Sichuan University, Chengdu, China. leizhang@scu.edu.cn}\\
\and
Philip S. Yu\thanks{Department of Computer Science, University of Illinois at Chicago, Chicago, USA. psyu@uic.edu}
\thanks{Institute for Data Science, Tsinghua University, Beijing, China.}
}
\date{}

\maketitle


\begin{abstract} \small\baselineskip=9pt
Recently, considerable attention has been devoted to the prediction problems arising from heterogeneous information networks. In this paper, we present a new prediction task, Neighbor Distribution Prediction (NDP), which aims at predicting the distribution of the labels on neighbors of a given node and is valuable for many different applications in heterogeneous information networks. The challenges of NDP mainly come from three aspects: the infinity of the state space of a neighbor distribution, the sparsity of available data, and how to fairly evaluate the predictions. To address these challenges, we first propose an Evolution Factor Model (EFM) for NDP, which utilizes two new structures proposed in this paper, i.e. Neighbor Distribution Vector (NDV) to represent the state of a given node's neighbors, and Neighbor Label Evolution Matrix (NLEM) to capture the dynamics of a neighbor distribution, respectively. We further propose a learning algorithm for Evolution Factor Model. To overcome the problem of data sparsity, the learning algorithm first clusters all the nodes and learns an NLEM for each cluster instead of for each node. For fairly evaluating the predicting results, we propose a new metric: Virtual Accuracy (VA), which takes into consideration both the absolute accuracy and the predictability of a node. Extensive experiments conducted on three real datasets from different domains validate the effectiveness of our proposed model EFM and metric VA.
\end{abstract}

\section{Introduction}
As part of the recent surge of research on information networks, considerable attention has been devoted to prediction problems in heterogeneous information networks. The existing researches, however, mainly focus just on the predictions around a single link. For example, some works are interested in predicting whether or when a link will be built in the future \cite{LPUSL:Hasan,CSLP:Leroy,NPAMILP:Lichtenwalter,LPMFLP:Wang,LPIRD:Taskar,WWIH:Sun}, and some works concern predicting strength of a link, such as predicting the ratings that customers will give to items or locations \cite{paterek:improving,savia:latent,Gorrell:generalized,LAPRUSG:Bao}. Existing researches surprisingly pay little attention to the prediction of neighbor distributions, where states of neighbors are considered as a whole.
\begin{figure}[!ht]
    \centering
    \epsfig{file=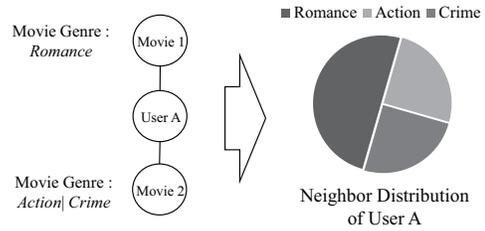, width=2.6in}
    \caption{Neighbors Distribution}
    \label{Figure:Transfer}
\end{figure}

Fig. \ref{Figure:Transfer} offers an illustration of neighbor distribution. The left part of Fig. \ref{Figure:Transfer} illustrates \textit{User A} has two movie nodes as its neighbors, which means \textit{User A} rented two movies. The neighbor distribution of node \textit{User A} is the distribution of labels on its neighbor nodes, as shown by the right part of Fig. \ref{Figure:Transfer}, where the neighbors of a given node have equal weight, and the weight of a neighbor node is uniformly divided by its labels.

The neighbor distribution of a node usually evolves over time. For example, a user might rent movies of different genres as his/her taste changes. Such evolution makes the prediction of neighbor distributions valuable for many different applications.

\textbf{Motivating Example} For an online sports video provider, the type distribution of subscribers is crucial to develop its sales strategy. The provider may be misled, if it only takes the recent sales data into consideration, and ignores the evolutionary feature of the type distribution. For example, the soccer fans are the major subscribers on May 2014, which may lead the provider to put more soccer advertisements online. However, the soccer fans are increasing slowly on May, and become the major subscribers on June 2014, as the opening of four-yearly soccer celebration "World Cup". Traditional recommender system methods may ignore the tiny increase of soccer fans on May.

In this paper, we aim at the problem of predicting the neighbor distribution of a given node in a heterogeneous information network, which has three main challenges we have to overcome:

$\bullet $ \textbf{Infinite state space of neighbor distributions}
Since the fraction of a label is a real value, the number of possible states of a neighbor distribution is theoretically infinite. Traditional temporal models such as Markov chain cannot serve our goal because they often assume a finite state space.

$\bullet $ \textbf{Sparsity of heterogeneous links}
In most cases, the links between one specific node and its heterogeneous neighbors are relatively sparse compared with the huge volume of a whole data set, e.g., "publishing" in DBLP, "rating" in Netflix and "checking in" in Foursquare. The sparsity of links between heterogeneous nodes makes it harder to mine sufficient meaningful patterns for individuals.

$\bullet $ \textbf{Fairly evaluating predictions}
Not all nodes are equally predictable, hence the traditional metrics that just take absolute accuracy into account are unable to appropriately assess the predictions for the nodes that are less predictable. We need a new metric that can treat every node fairly.

In this paper, inspired by the idea of Factor Model \cite{URL:webb, koren:matrix}, we propose an Evolution Factor Model (EFM) to accurately predict neighbor distributions from the sparse data.
Our main contributions can be summarized as follows:

\begin {enumerate} [(1)]
\item We introduce an Evolution Factor Model (EFM) for accurately predicting neighbor distributions. EFM employs our proposed data structures, Neighbor Distribution Vector (NDV) and Neighbor Label Evolution Matrix (NLEM), to represent the infinite state space of a neighbor distribution and capture the evolution of neighbor distributions respectively.

\item We propose a new prediction metric, Virtual Accuracy (VA), which takes into consideration both the absolute accuracy and the difficulty of a prediction to fairly evaluate the prediction results of nodes with different predictabilities.

\item We conduct extensive experiments on three real datasets, and compare EFM with an empirical method and two existing methods. The results validate the performance of our proposed model, algorithm and accuracy metric.
\end{enumerate}

The rest of this paper is organized as follows. We give the problem definition and formalization in Section 2. In Section 3, we describe our prediction model EFM, and further present the learning algorithm for EFM. We discuss the predictability of nodes and propose a prediction metric in Section 4. We present the experimental results and analysis in Section 5. Finally, we discuss related works in Section 6, and conclude in Section 7.

\section{Problem Definition}
\subsection{Heterogeneous Information Network}

A heterogeneous information network contains multiple types of objects and links. In this paper, we only consider those heterogeneous information networks with star network schema \cite{CEMTODSN:Sun}, i.e., links only exist between the center type of nodes as \textbf{target nodes}, and several other types of nodes as \textbf{attribute nodes}. For example, in Location Based Social Network, the target nodes are users, and the attribute nodes can be venues, ratings or tips.

We denote the heterogeneous information networks with star network schema by $G={\left \langle V,E \right \rangle}$, where $V$ is the node set and $E$ is the link set. We denote the target and attribute node set by $\mathcal{X} \subset V$ and $\mathcal{U} \subset V$ respectively. The nodes and the links in networks are being constructed and destructed over time. In order to capture the dynamics, we use time window, which is denoted by $T$, to capture the neighbor distribution with timeliness from dynamic networks. The node set and the attribute node set in $T$ are denoted by $V_T$ and $\mathcal{U}_T$. We use $T_h$, $T_c$ and $T_f$ to represent the historical, current and future time windows respectively.

\subsection{Label Distribution Vector}
$\\$
Assuming the universal label set of a given attribute node set $\mathcal{U}$ is denoted by $\beta_{\mathcal{U}} =\left \{ \beta^{(1)}_{\mathcal{U}},...,\beta^{(i)}_{\mathcal{U}},...,\beta^{(n)}_{\mathcal{U}} \right \}$, where $\beta^{(i)}_{\mathcal{U}}$ is a label, and $n$ is the number of label types, the definition of label distribution vector is given as follow:

\begin{Definition}{
    \rm
    \textbf{Label Distribution Vector (LDV)} For a given attribute node $u \in \mathcal{U}$, its LDV, $\overrightarrow{v}_u \in \mathbb{R}^{n}$, is defined as:
    \begin{eqnarray}
        \overrightarrow{v}_u = (v_{u}^{(1)},...,v_{u}^{(i)},...,v_{u}^{(n)}),
    \end{eqnarray}
    where $n$ is the number of all attribute nodes' label types;  $v_{u}^{(i)}=\frac{I_{u}(\beta^{(i)}_{\mathcal{U}})}{\sum_{k=1}^{n}I_{u}(\beta^{(k)}_{\mathcal{U}})}$,
    $I_{u}(\beta^{(i)}_{\mathcal{U}})$ is 1 if $u$ has the label $\beta^{(i)}_{\mathcal{U}}$, and 0, otherwise.
}\end{Definition}

LDV measures the label distribution of an attribute node, which is fixed. For example, the labels of an article are subject areas, thus the LDV depicts the direction of this article.

\subsection{Neighbor Distribution Vector}

\begin{Definition}{
    \rm
    \textbf{Neighbor Distribution Vector (NDV)}
    For a given target node $x \in \mathcal{X}$, the NDV of $x$'s attribute node neighbors $\mathcal{U}'_T \subseteq \mathcal{U}_T $ in time window $T$, $ \overrightarrow{w}_{x}(\mathcal{U}'_T) \in \mathbb{R}^{n}$, is defined as:
    \begin{eqnarray}
        \overrightarrow{w}_x(\mathcal{U}'_T) =
        (w_{x}^{(1)},...,w_{x}^{(n)}),
    \end{eqnarray}
    where
    $n$ is the number of the given attribute nodes' label types; $w_{x}^{(i)}=\frac{\sum_{u \in \mathcal{U}'_T}(v_{u}^{(i)}) + 1 }{\left | \mathcal{U}'_T \right | + n}, i \in [1,n]$.
}\end{Definition}

Hereinafter, we just denote a NDV by $\overrightarrow{w}_x$ if the context is unambiguous. Note that: (1) For smoothing, we add 1 in the numerator and $n$ in the denominator of $w_{x}^{(i)}$. (2) A node has an NDV corresponding to each different type of attribute node neighbors. For example, in DBLP, there are two type of attribute nodes, which are "Articles" and "Journals"; therefore, the target node "Scholar" should have two NDVs, one for its "Article" neighbors and the other for its "Journal" neighbors.

\subsection{Problem Statement}
$\\$
Based on NDV, we can formally state the problem of Neighbor Distribution Prediction (NDP) as follow :

Assigned the historical, current and future time window $T_h$, $T_c$ and $T_f$ respectively, given a target node $x$ and its NDV
of $x$'s attribute node neighbors $\mathcal{U}$ in time window $T_h$ and $T_c$, we want to predict $\overrightarrow{w}_x(\mathcal{U}'_{T_f})$.

\section{Evolution Factor Model}
In this section, we describe our Evolution Factor Model (EFM). At first, we briefly introduce the basic idea of Factor Model.

\subsection{Evolution Factor Model}
$\\$
Recommender system methods based on latent factor matrix model take the data in historical and current time window as a whole, but dismiss the evolution of the network. In order to capture the dynamics, Evolution Factor Model first stores the probability of changes from one label to another, which leads to the following definition of Neighbor Label Evolution Matrix (NLEM).

\begin{Definition}{
    \rm
    \textbf{Neighbor Label Evolution Matrix}
    For a given node $x$, its Neighbor Label Evolution Matrix of attribute nodes $\mathcal{U}$ from time window $T_p$ to $T_q$, denoted by $L^{<\mathcal{U}'_{T_p}, \mathcal{U}'_{T_q}>}_x \in \mathbb{R}^{n \times n}$, is a matrix in which a cell $L_x(i,j)$ is the probability that $x$'s neighbor label changes from $\beta^{(j)}$ to $\beta^{(i)}$, i.e.,
    \begin{eqnarray}
        L_x(i,j) = P(\beta^{(i)}|\beta^{(j)}).
    \end{eqnarray}
}\end{Definition}

\textbf{Evolution Factor Model} Based on NLEM, for a given target node $x$, we can predict its NDV $\overrightarrow{w}_x(\mathcal{U}'_{T_f})$ as the transformation of its historical NDV, $\overrightarrow{w}_x(\mathcal{U}'_{T_h + T_c})$ through its NLEM $L^{<\mathcal{U}'_{T_h}, \mathcal{U}'_{T_c}>}_x$, which leads to our EFM as follows:

\begin{eqnarray}
    & \overrightarrow{w}_x(\mathcal{U}'_{T_f}) =
    L^{<\mathcal{U}'_{T_h}, \mathcal{U}'_{T_c}>}_x \times \overrightarrow{w}_x(\mathcal{U}'_{T_h+T_c}).
    \label{Evolution_Factor_Model}
\end{eqnarray}

\begin{figure}[!h]
    \centering
    \epsfig{file=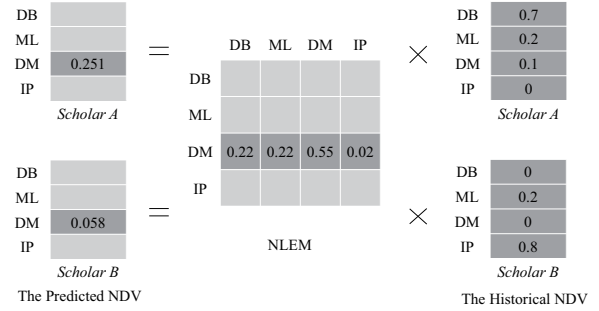, width=3in}
    \caption{An example of Evolution Factor Model}
    \label{Figure:Evolution}
\end{figure}
Fig. \ref{Figure:Evolution} gives an example that shows how EFM works. In Fig. \ref{Figure:Evolution}, there are four labels representing different research directions: DB (Data Base), ML (Machine Learning), DM (Data Mining), IP (Image Processing). A cell $(i, j)$ of NLEM represents the probability that scholars change their research directions from $i$ to $j$. By given an NLEM learned from historical data, and the current NDVs of \textit{Scholar A} and \textit{Scholar B}, we can infer their next NDVs by transforming their historical NDVs by the learned NLEM.

Note that the essence of the matrix product in EFM is different from that in a general factor model. In a general factor model, the matrix product is static, which consider the data in historical and current time window as a whole. In contrast, NLEM, in our EFM, which can capture the changes from historical time window to current time window agilely, and changes as the given target node's neighbor labels evolve over time.

\subsection{Model Learning}
$\\$
To overcome the issue of data sparsity, in the light of the heuristic knowledge that similar individuals have similar behaviors, we first apply a clustering algorithm to all the target nodes, and learn the NLEM for the cluster which $x$ belongs to. According to this idea, given target node $x$ and the node set $\mathcal{X}'$ consisting of the nodes belonging to the same cluster of $x$, we can learn NLEM as follow:
\begin{eqnarray}
    L^{<\mathcal{U}'_{T_h}, \mathcal{U}'_{T_c}>}_x = \underset{L}{argmin} \sum_{x' \in \mathcal{X}'}{} \epsilon^2,
    \label{minimzing problem}
\end{eqnarray}
where $\epsilon = |  { \widehat{\overrightarrow{w}}_{x'}(\mathcal{U}'_{T_c})} -  \overrightarrow{w}_{x'}(\mathcal{U}'_{T_c}) |$, $L \in \mathbb{R}^{n \times n}$, $\widehat{\overrightarrow{w}}_{x'}(\mathcal{U}'_{T_c}) = L \times \overrightarrow{w}_{x'}(\mathcal{U}'_{T_h})$.

Note that $\widehat{\overrightarrow{w}}_{x'}(\mathcal{U}'_{T_c})$ is an estimate of $\overrightarrow{w}_{x'}(\mathcal{U}'_{T_c})$ for target node $x'$.
So NLEM is actually defined as the optimal matrix that minimizes the overall error of the estimates over all target nodes in $\mathcal{X}$. Thus, for learning NLEM, we adopt least square method to establish linear regression, indicating the NDVs in $T_h$ and $T_c$ as follows:
\begin{eqnarray}
    L^{<\mathcal{U}'_{T_h}, \mathcal{U}'_{T_c}>}_x =(X^T X)^{-1} X^T Y,
    \label{LS formula}
\end{eqnarray}
where
\begin{center}
    $X=
    \begin{Bmatrix}
        X^{(1)}=\overrightarrow{w}_{x^{(1)}}(\mathcal{U}'_{T_h})\\
        ...\\
        X^{(N)}=\overrightarrow{w}_{x^{(N)}}(\mathcal{U}'_{T_h})
    \end{Bmatrix},$\\

    $Y=
    \begin{Bmatrix}
        Y^{(1)}=\overrightarrow{w}_{x^{(1)}}(\mathcal{U}'_{T_c})\\
        ...\\
        Y^{(N)}=\overrightarrow{w}_{x^{(N)}}(\mathcal{U}'_{T_c})
    \end{Bmatrix}.$
\end{center}
The learning algorithm for EFM is shown in Algorithm 1.
\begin{algorithm}[!h]
  \caption{ \emph{Learning Algorithm for EFM $(x, T_h, T_c, \mathcal{X}_{T_h}, \mathcal{X}_{T_c}, \mathcal{U}, K )$} }
  \begin{algorithmic}[1]
    \REQUIRE ~~ \\
        $x$: A given node;\\
        $T_h$: Assigned historical time window;\\
        $T_c$: Assigned current time window;\\
        $\mathcal{X}_{T_h}$: A subset of $\mathcal{X}$ in time window $T_h$;\\
        $\mathcal{X}_{T_c}$: A subset of $\mathcal{X}$ in time window $T_c$;\\
        $\mathcal{U}$: The given attribute node set;\\
        $K$: The parameter of $K$-means;

    \ENSURE ~~ \\
       $L^{<\mathcal{U}'_{T_h},\mathcal{U}'_{T_c}>}_x$: The NLEM needed to be learned;

    \STATE $\mathcal{X'}=\Phi$, $X =\textbf{0}$, $Y =\textbf{0}$, $\mathrm{W}=\{x'\:|\:x' \in \mathcal{X}_{T_h}, \mathcal{X}_{T_c}$\};
    \FOR{each $x' \in \mathrm{W}$}
        \STATE \textbf{Compute} $\overrightarrow{w}_{x'}(\mathcal{U}'_{T_h})$ and $\overrightarrow{w}_{x'}(\mathcal{U}'_{T_c})$;
    \ENDFOR
    \STATE \textbf{Do} $K-$means on $\mathrm{W}$ based on the similarities between $\overrightarrow{w}_{x'}(\mathcal{U}'_{T_h})$;
    \STATE $\mathcal{X'} = $ \{the nodes of the cluster that $x$ belongs to\};
    \FOR{$i = 1 $ to $|\mathcal{X'}|$}
        \STATE $X^{(i)}=\overrightarrow{w}_{x^{(i)}}(\mathcal{U}'_{T_h})$;
        \STATE $Y^{(i)}=\overrightarrow{w}_{x^{(i)}}(\mathcal{U}'_{T_c})$;
    \ENDFOR
    \STATE \textbf{Compute} $L^{<\mathcal{U}'_{T_h},\mathcal{U}'_{T_c}>}_x$ according to Equation (\ref{LS formula}), where inverse matrix is computed by Gauss Jordan method;
  \end{algorithmic}
\end{algorithm}

In our learning algorithm, any classical clustering algorithm is qualified for our learning algorithm. We choose $K$-means as the clustering algorithm, where the similarities are measured by Euclidean distances between NDVs. The selection of $K$ is discussed in Section V.

\subsection{Prediction}
$\\$
Given a target node $x$, its NDV $\overrightarrow{w}_x(\mathcal{U}'_{T_h + T_c})$ and the learned NLEM $L^{<\mathcal{U}'_{T_h}, \mathcal{U}'_{T_c}>}_x$, the prediction of $\overrightarrow{w}_x(\mathcal{U}'_{T_f})$ can be made by EFM(Equation (\ref{Evolution_Factor_Model})).

\section{Prediction Metric}

\subsection{Normalized Absolute Accuracy}
$\\$
We can intuitively measure the absolute accuracy of predictions for the given node $x$ in terms of the Euclidean distance between a true $\overrightarrow{w}_x$ and its estimate $\widehat{\overrightarrow{w}}_x$.

By Definition 2, components of an NDV are positive and the sum of them is equal to 1, so the Euclidean distance between any two NDVs is less than or equal to $\sqrt{2}$. Then we can define the normalised absolute accuracy as follow:
\begin{eqnarray}
    &\eta_x = 1 - \frac{d(\widehat{\overrightarrow{w}}_x, \overrightarrow{w}_x)}{\sqrt{2}},
\end{eqnarray}
where $d(\widehat{\overrightarrow{w}}_x, \overrightarrow{w}_x)$ is the Euclidean distance between $\widehat{\overrightarrow{w}}_x$ and $\overrightarrow{w}_x$.

\subsection{Predictability}
$\\$
As we have mentioned, it is unfair to assess a prediction just in terms of absolute accuracy, since the predictability of nodes are different.

Intuitively, the predictability of a node is relevant to its susceptibility to the similar homogeneous nodes. Specifically, the easier the node can be influenced by others, the more disordered its temporal pattern is, and the greater its predictability is. For example, in a given research field, the leading scholars' directions are difficult to capture, because they change their research directions rarely and such changes are mainly breakthroughs. These changes can hardly be predicted compared with their long-term stable studies. In contrast, the research direction of a PhD candidate is more likely influenced by his/her supervisor or the leading scholars. Inspired by this observation, we can define Prediction Difficulty as the measure on how difficult to predict a given node's NDV.

\begin{Definition}{
    \rm
    \textbf{Prediction Difficulty (PD)}
    For a node $x$, the prediction difficulty of its NDV of attribute node neighbors $\mathcal{U}'_T$ in time window $T$, denoted by $g_x(\mathcal{U}'_T)$ is defined as:
    \begin{eqnarray*}
        &g_x(\mathcal{U}'_T) = 1 - h_x(\mathcal{U}'_T)/2,
    \end{eqnarray*}
    where $h_x(\mathcal{U}'_T)$ is the temporal entropy of $x$'s NDV in time window $T$, and
    \begin{eqnarray*}
        &h_x(\mathcal{U}'_T) = -\sum_{i=1}^{n}w_{x}^{(i)}(\mathcal{U}'_T) log_n w_{x}^{(i)}(\mathcal{U}'_T).
    \end{eqnarray*}
}\end{Definition}

Note that, $w^{(i)} \in (0,1),$ so $h_x(\mathcal{U}'_T) > 0$, and when $w^{(i)}= 1/n, \forall i \in [1, n]$, $h_x(\mathcal{U}'_T)$ reaches the maximum, which equals $-\sum_{i=1}^{n} \frac{1}{n}log_n\frac{1}{n}=1$. Thus, $h_x(\mathcal{U}'_T) \in (0,1]$ and $g_x(\mathcal{U}'_T) \in [1/2,1)$. In Definition 4, we use temporal entropy $h_x(\mathcal{U}'_T)$ to measure how disordered a node's temporal pattern is. We can see the more disordered the temporal pattern, the greater the temporal entropy, and consequently the less the prediction difficulty, which is in line with our expectation.

\begin{table*}[!ht]\small
    \begin{center}
        \begin{tabular}{c|ccccc}
            \hline
                Dataset & Network & The type of     & The type of                & \#Neighbor's lables & \#Predicted Nodes\\
                        &         & predicted nodes & predicted node's neighbors &                     &\\
            \hline
            \hline
                DBLP & heterogeneous & author & paper & 25 & 1000\\
            \hline
                Netflix & heterogeneous & user & movie & 28 & 1000\\
            \hline
                Foursquare & heterogeneous & user & venue & 8 & 500\\
            \hline
        \end{tabular}
    \end{center}
    \caption{Summary of Datasets.}
    \label{table:Datasets}
\end{table*}
\subsection{Virtual Accuracy}
$\\$
Now we further define Virtual Accuracy based on absolute accuracy and prediction difficulty as follow:
\begin{Definition}{
    \rm
    \textbf{Virtual Accuracy (VA)}
    For a prediction of $\overrightarrow{w}_x(\mathcal{U},T_f)$,
    its Virtual Accuracy, denoted by $\delta_x$, is defined as:
    \begin{eqnarray}
        &\delta_x  = \eta_x \times g_x,
        \label{VA:formula}
    \end{eqnarray}
    where $\eta_x$ is the absolute accuracy and $g_x$ is the prediction difficulty.
}\end{Definition}

As Equation (\ref{VA:formula}) shows, we define VA of a prediction as the product of the absolute accuracy and the predictability of that prediction. Since $\eta_x$ and $g_x$ are both nonnegative, it is obvious that VA favors the predictions whose absolute accuracy and difficulty are both great. As we can see in later experiments, $\eta_x$ is negatively correlated with $g_x$. Thus even the absolute accuracy of a difficult prediction is low, the VA of it can still be expected be not low since its prediction difficulty is large. On the other hand, even the absolute accuracy of an easy prediction is high, the VA of it is expected to be low due to its small prediction difficulty.

\section{Experimental Evaluation}

\subsection{Datasets}
$\\$
We learn the NLEM from the NDVs in $T_h$ and $T_c$. For predicting neighbor distribution, we take the NDVs in $T_h + T_c$ as training set, and the NDVs in $T_f$ as test set.

The datasets we use to validate our model and algorithm are from three different domains, DBLP (a Coauthor Network), Netflix (a Movie Rental Network), and Foursquare (a Location Based Social Network). The summary of datasets is shown in Table \ref{table:Datasets}

\textbf{DBLP} \cite{URL:DBLP} indexes more than about $230$ million articles and contains massive links to home pages of computer scientists. The labels of "Article" contain $25$ directions on Computer Science, thus an NDV of a "Scholar" node consists of $25$ components. By assigned historical, current and future time window $T_h = [2006, 2010)$, $T_c = [2010, 2011)$ and $T_f = [2011, 2012]$, we randomly select $1000$ scholars who published articles in all the three time window.

\textbf{Netflix} \cite{URL:Netflix} contains about more than $100$ million rating records from about $480,000$ customers over about $17,000$ movie titles. The labels of "Movie" contain $28$ genres crawled from the website IMDb \cite{URL:IMDb}, thus an NDV of a "User" node consists of $28$ components. By assigned historical, current and future time window $T_h$ = [Apr.$12^{th}$ 2005, Oct.$12^{th}$ 2005), $T_c$ = [Oct.$12^{th}$ 2005, Nov.$12^{th}$ 2005) and $T_f$ = [Nov.$12^{th}$ 2005, Dec.$12^{th}$ 2005], we randomly select $1,000$ users who have movie rating records in all the three time window.

\textbf{Foursquare}\cite{LAPRUSG:Bao} involves about $4.3$ million friendships and about $80,000$ check-in tips of users during $3$ years. The labels of "Venue" contain $8$ categories given by Foursquare. By assigned historical, current and future time window $T_h$ = [$0^{^{th}}$ day, $966^{^{th}}$ day), $T_c$ = [$966^{^{th}}$ day, $996^{^{th}}$ day) and $T_f$ = [$996^{^{th}}$ day, $1026^{^{th}}$ day], we randomly select $500$ users who have check-in records during in all the three time window.

\subsection{Baseline}
$\\$
In order to demonstrate the effectiveness of our EFM, we compare our method with the following baseline methods:

\begin {enumerate} [$\bullet$]
\item \textbf{MVM} (Mean Value Method) MVM is an empirical method. It takes the mean of the latest NDVs of the nodes in the cluster that the predicted node belongs to, as the estimate of the next NDV of a given node.

\item \textbf{MF} (Basic Matrix Factorization)\cite{URL:webb} MF is proposed by B. Webb to solve the movie recommender problem in Netflix Price. MF assumes the features of objects can be expressed as a series of factors, and different types of objects have factors with the same amount. When predicting the preference of the given objects of type $A$ for the objects of type $B$, the preferences (which is called "ratings" in many cases) can be expressed as the product of the factors of the given objects of type $A$ and $B$. The general expression of factor model is:
    \begin{eqnarray}
        &R = PQ^T,
        \label{General_Factor_Model}
    \end{eqnarray}
    where $P \in \mathbb{R}^{N \times D}$ is the factor matrix of the objects of type $A$. $Q \in \mathbb{R}^{M \times D}$ is the factor matrix of objects of type $B$. $N$ and $M$ are the number of the objects of type A and the number of the objects of type B, respectively. $D$ is the factor number.

\item \textbf{BiasedMF} (Biased Matrix Factorization)\cite{paterek:improving} BiasedMF is proposed by Paterek, which is an extension of Basic Matrix Factorization. BiasedMF adds biased rates to the objects of either type. The prediction formula is:
    \begin{eqnarray}
        &\hat{r}_{u,m}=b_u+b_m+\sum_{k=1}^{n}p_{u,k} \cdot q_{m,k},
        \label{BiasedMF:Formula}
    \end{eqnarray}
    where $\hat{r}_{u,m}$ is an estimate of rate that the object $u$ of type $A$ gives to the object $m$ of type $B$. The $p_{u,k}$ and $q_{m,k}$ are the cells of the factor matrixes of type $A$ and type $B$ respectively. $b_u$ and $b_m$ are the biases of object $u$ and $m$ respectively.

\end {enumerate}

In our experiments, we set the parameters of MF and BiasedMF as learning rate $\eta = 0.001$ and punishing parameter $\lambda = 0.02$, as suggested by Paterek \cite{paterek:improving} and Gorrell et al. \cite{Gorrell:generalized}. We choose the number of NDV components as the latent feature numbers in MF and BiasedMF. Thus the feature numbers of DBLP, Netfilix and Foursquare are $25, 28$ and $8$ respectively.

\subsection{The Determination of $K$}
\begin{figure}[!ht]
    \begin{minipage}[t]{0.25\textwidth}
         \subfigure[DBLP and Netflix]{\epsfig{file=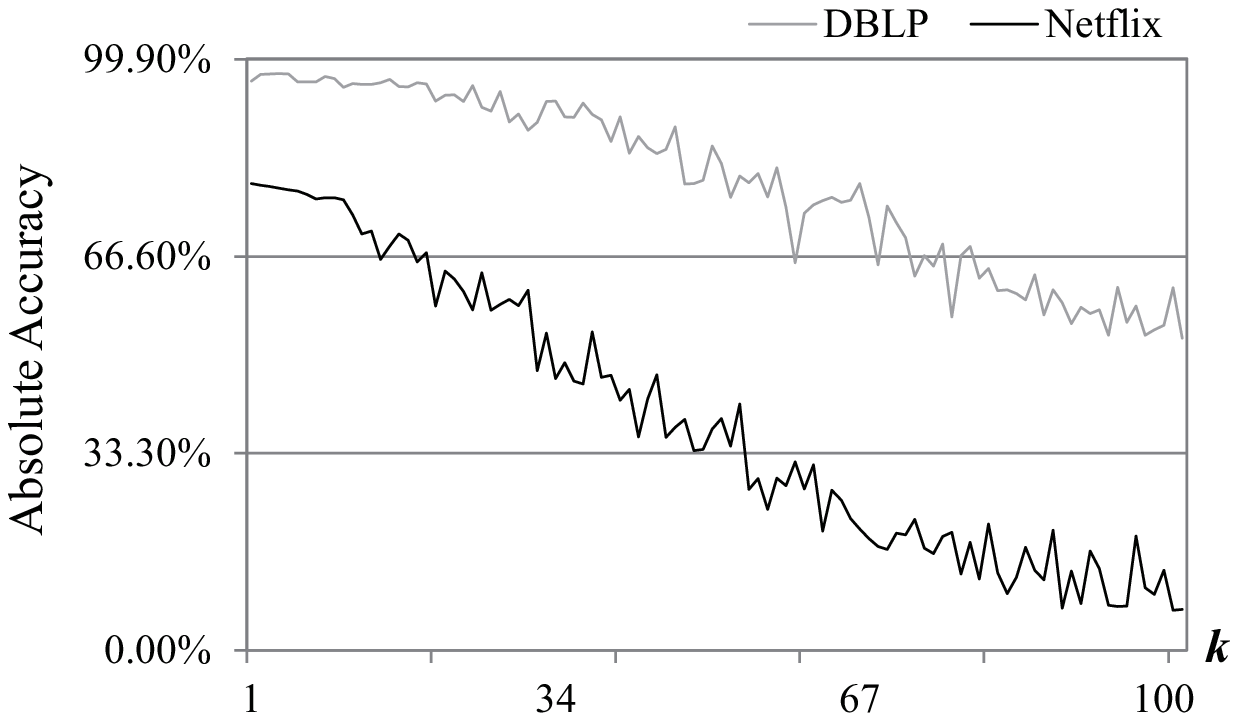, width=1.6in}}
    \end{minipage}%
    \begin{minipage}[t]{0.25\textwidth}
         \subfigure[Foursquare]{\epsfig{file=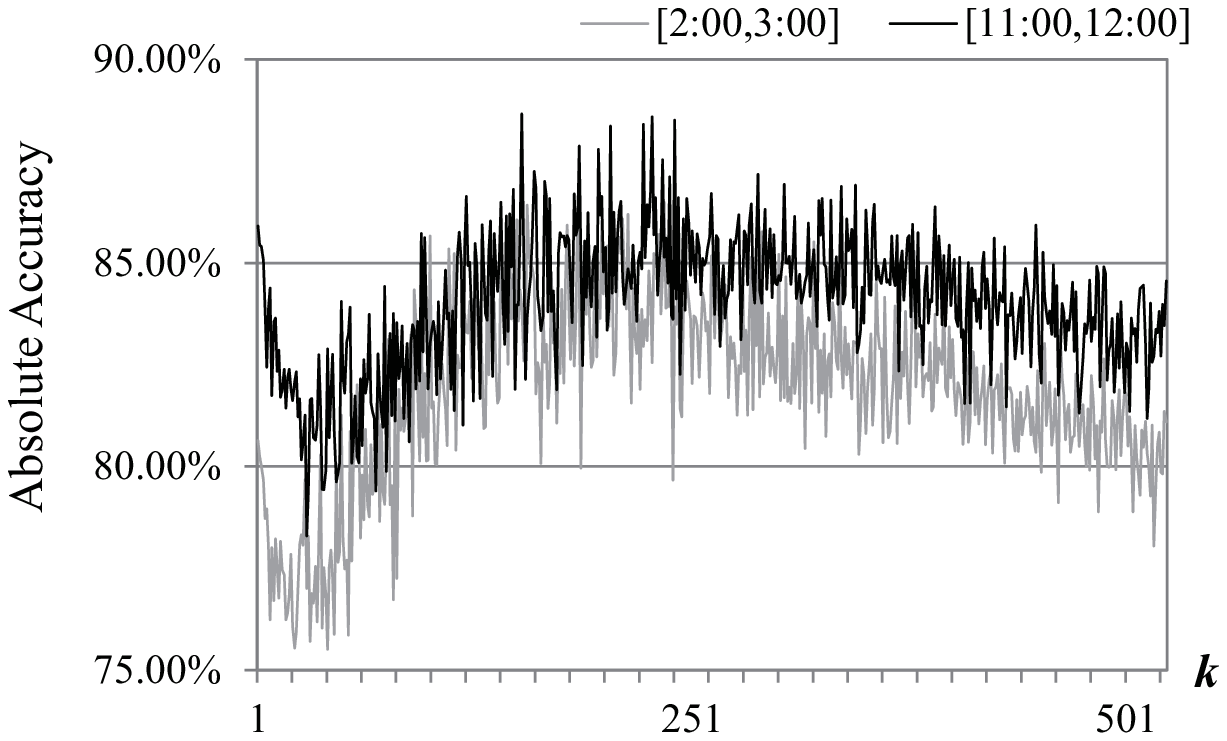, width=1.6in}}
    \end{minipage}%
    \caption{The Selection of $K$}
    \label{Figure:The_Selection_of_K}
\end{figure}

$\\$
Learning Algorithm for EFM requires the number of clusters, $K$, as the input when it invokes a $K$-means procedure, so we have to determine $K$ before we start our experiments. For each dataset, we first randomly select $500$ nodes from it, then apply our model to make predictions for these nodes and choose the $K$ that maximizes the average absolute accuracy of the predictions. As Fig. \ref{Figure:The_Selection_of_K} shows, we finally get $K = 5$ for DBLP, $K=1$ for Netflix, and $K =155, 156$ for Foursquare during $[2:00, 3:00]$ and $[11:00, 12:00]$ respectively.

\subsection{The Validation of Predictability}
$\\$
Now we investigate how the absolute accuracy of a prediction correlates with its prediction difficulty. For each dataset, we first rank the nodes by PD in descending order, then divide the nodes into five groups. The nodes in a same group have equal PD. Finally we observe the absolute accuracies by applying EFM and three baseline methods, MVM, MF and BiasedMF, on the five groups respectively.

\begin{figure}[!ht]
    \centering
    \epsfig{file=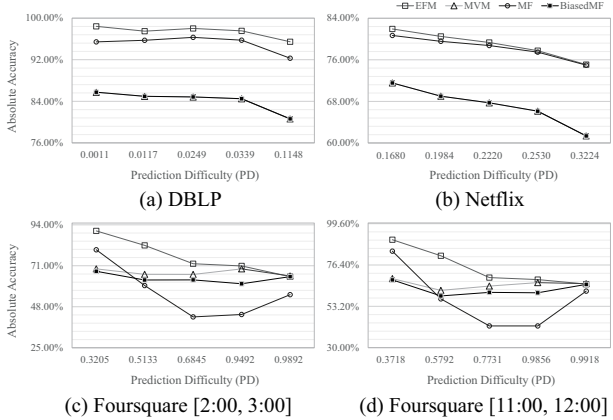, width=3.2in}
    \caption{The Relationship between Absolute Accuracy and Prediction Difficulty on Three Datasets}
    \label{Figure:TemporalEntropies_DBLP_Netflix}
\end{figure}

The results are shown in Fig. \ref{Figure:TemporalEntropies_DBLP_Netflix}. We can see that the absolute accuracies of the methods we use in the experiments decrease in overall with the increase of the prediction difficulty. Such result validates our assumption that the more disordered the temporal pattern of a node is, the greater its predictability is. It also shows the necessity to assess a model by a fair metric which should take the predictability into consideration.

Note that, on Foursquare, the absolute accuracies of baseline methods do not decrease linearly with PD. It is because the human's daily routines are not all the same, which leads to the fluctuation of absolute accuracies of baseline methods. The absolute accuracy of EFM, however, has a linear decrease with PD. It is because EFM is not limited to the recognition of daily pattern, but instead takes the evolution regularity (represented by the neighbor label evolution matrix in EFM) into consideration. For example, the office workers who like nightlife can go to the nightclub for sleepover only on Weekends. For the nodes of that type, the two empirical methods can not perform as expected because the activities of sleepover on weekends are not common to everyone (which is the reason why MVM's curve fluctuates), or to an individual on everyday (which is the reason why the curve of MF and BiasedMF fluctuate).

\subsection{The Comparison between EFM and Baseline Methods}
$\\$
\begin{figure}[!ht]
    \centering
    \epsfig{file=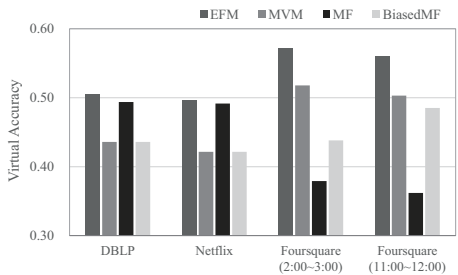, width=2.4in}
    \caption{The Comparison of VA between EFM and Baseline Methods}
    \label{Figure:VA_Comparison}
\end{figure}
In this part, we compare the virtual accuracy of EFM with three baseline methods: MVM, MF and BiasedMF. The summarized result is shown in Fig. \ref{Figure:VA_Comparison} and the detailed result is listed in Table \ref{Table:VA}. Our remarks on the result are as follows:

\begin{table}[!ht]
    \scriptsize
    \begin{center}
        \begin{tabular}{ccccc}
            \hline
                       &      &         & Foursquare   & Foursquare\\
                Method & DBLP & Netflix & [2:00, 3:00] & [11:00, 12:00]\\
            \hline
            \hline
                EFM & \textbf{0.5049} & \textbf{0.4960} &	\textbf{0.5722} & \textbf{0.5606}\\
            \hline
                MVM & 0.4359 & 0.4216 & 0.5179 & 0.5032\\
            \hline
                MF & 0.4937 & 0.4917 & 0.3793 & 0.3619\\
            \hline
                BiasedMF & 0.4360 & 0.4217 & 0.4381 & 0.4852\\
            \hline
        \end{tabular}
    \end{center}
    \caption{Virtual Accuracies of EFM, MVM, MF and BiasedMF}
    \label{Table:VA}
\end{table}

\begin {enumerate} [(1)]

\item EFM performs far better than MVM on all the datasets, while MVM has the worst performance on the Netflix and DBLP datasets.

\item Although having the worst performance on the two Foursquare datasets, MF does have a good performance on Netflix dataset comparing with the other baseline methods, since MF is originally proposed for the movie recommendation problem in Netflix. EFM, however, still performs better than MF, which is because EFM can take into consideration not only the profile of the predicted nodes, but also the evolution regularity.

\item As shown in Fig. \ref{Figure:VA_Comparison},
     EFM outperforms BiasedMF, which performs similarly to MVM on DBLP and Netflix, but far worse on Foursquare.

\item EFM outperforms MVM, MF and BiasedMF especially on Foursquare. This is because the three baseline methods only pay attention to the daily pattern (MVM) or the profile of users (MF and BiasedMF). However, the activities in Foursquare are limited to not only the daily pattern or profile of users, but also the evolution regularity over weeks, even months.

\end {enumerate}
In summary, EFM is a robust and effective method. The VA of EFM is generally better than all the baseline methods.

\section{Related Work}
$\\$
Three domains are relevant to our work, namely link prediction, rating prediction and factor model.

\textbf{Link Prediction}: Hasan et al. \cite{LPUSL:Hasan} first introduces supervised learning to predict whether a link will be built in the future. Wang et al. \cite{LPMFLP:Wang} introduces probabilistic model for link prediction. Leroy et al. \cite{CSLP:Leroy} solves the cold start problem in link prediction. Lichtenwalter et al. \cite{NPAMILP:Lichtenwalter} proposes new perspectives and methods in link prediction. Taskar et al. \cite{LPIRD:Taskar} propose a method to address the problem of link prediction in heterogeneous networks, based on the observations of the attributes of the objects. Sun et al. \cite{WWIH:Sun} extends the traditional link prediction to relationship prediction, which not only predicts whether it will happen, but also infers when it will happen. However, Sun's work, still focuses on the predictions of a single link.

\textbf{Rating Prediction (Recommender System)}: Basically, the methods predicting ratings of links fall into two categories: memory-based algorithms and model-based algorithms. The memory-based algorithms directly make predictions based on homogeneous neighbors of a given node \cite{Linden:Amazon,Sarwar:Item-based,Yu:Probabilistic}, while model-based algorithms make predictions based on a prediction model learned in advance. Savia et al.\cite{savia:latent} proposes a prediction model based on bayesian networks. These existing methods pay insufficient attention to the evolution of neighbor distributions.

\textbf{Factor Model}: Factor Model assumes the features of objects can be expressed as a series of factors, which is also called Matrix Factorization (MF). MF is first proposed by Webb\cite{URL:webb} to solve the movie recommender problem in Netflix Price. Based on Webb's work, G. Gorrell et al.\cite{Gorrell:generalized} optimize the learning rate and the punishing parameter in MF. Paterek \cite{paterek:improving} proposes Biased Matrix Factorization (BiasedMF) to improve the performance of MF. However, the existing methods can not capture the dynamics of neighbor distributions agilely, which is exactly why we propose a new model EFM for our goal.

\section{Conclusion}
In this paper, we present a new prediction problem, Neighbor Distribution Prediction in heterogeneous information network. To address this problem, we propose an Evolution Factor Model (EFM), which takes Neighbor Label Evolution Matrix (NLEM) as the dynamic factor, and predicts the next NDV of a given node by transforming its current NDV by the NLEM. We also propose a learning algorithm for EFM, which learns the NLEM from the homogeneous nodes which are in the same cluster as a given node.

For fairly evaluating the predictions made by different methods, we propose Virtual Accuracy, which not only measures the absolute accuracy, but also takes the difficulty of a prediction into consideration.

We conduct the experiments on the datasets from three different applications, and compare EFM with three baseline methods: Mean Value Method, Basic Matrix Factorization and Biased Matrix Factorization. The results show EFM outperforms all the baseline methods in overall.

\section*{Acknowledgments}
This work is supported by the National Science Foundation of China under Grant Nos. 61173099, 61103043 and the Doctoral Fund of Ministry of Education of China under Grant No. 20110181120062. This work is also supported in part by NSF through grants CNS-1115234, DBI-0960443, and OISE-1129076, and US Department of Army through grant W911NF-12-1-0066.

\bibliographystyle{abbrv}

\bibliography{ltexpprt}

\end{document}